\documentclass[prl,twocolumn,showpacs,superscriptaddress,floatfix]{revtex4}
 \def\LineNumbers{false}	
 \def\AuthorList{true} 

\usepackage{ifthen}
\usepackage{lineno}
\usepackage{slashed}
\usepackage{epsfig}
\usepackage{amsmath}
\usepackage{comment}
\usepackage{longtable}
\usepackage{subfigure}
\usepackage{color}

\newcommand{\PYTHIA}     {{\sc pythia}}
\newcommand{\ppbar}{$p\overline{p}$}
\newcommand{\unit}[1]{\,\mathrm{#1}}  
\newcommand{\gevcc}{GeV/$c^2$}
\def\missET {{\not\!\!E_T}}
\def\missETvec {{\not\!\!\vec{E}_T}}

\newcommand{\zhllbb}{$ZH\rightarrow \ell^+\ell^-b\overline{b}$}
\newcommand{\zll}{$Z\rightarrow\ell^{+}\ell^{-}$}
\newcommand{\hbb}{$H\rightarrow b \bar{b}$}
\newcommand{\zee}{$Z\rightarrow e^{+}e^{-}$}
\newcommand{\zmm}{$Z\rightarrow \mu^{+}\mu^{-}$}

\begin{document}

\ifthenelse{\equal{\LineNumbers}{true}}{
 \pagewiselinenumbers
 \linenumbers
}


\title{Search for the standard model Higgs boson decaying to a $b\bar{b}$ pair in  
events with two oppositely-charged leptons using the full CDF data set}
\ifthenelse{\equal{\AuthorList}{true}}{
  \affiliation{Institute of Physics, Academia Sinica, Taipei, Taiwan 11529, Republic of China}
\affiliation{Argonne National Laboratory, Argonne, Illinois 60439, USA}
\affiliation{University of Athens, 157 71 Athens, Greece}
\affiliation{Institut de Fisica d'Altes Energies, ICREA, Universitat Autonoma de Barcelona, E-08193, Bellaterra (Barcelona), Spain}
\affiliation{Baylor University, Waco, Texas 76798, USA}
\affiliation{Istituto Nazionale di Fisica Nucleare Bologna, $^{ee}$University of Bologna, I-40127 Bologna, Italy}
\affiliation{University of California, Davis, Davis, California 95616, USA}
\affiliation{University of California, Los Angeles, Los Angeles, California 90024, USA}
\affiliation{Instituto de Fisica de Cantabria, CSIC-University of Cantabria, 39005 Santander, Spain}
\affiliation{Carnegie Mellon University, Pittsburgh, Pennsylvania 15213, USA}
\affiliation{Enrico Fermi Institute, University of Chicago, Chicago, Illinois 60637, USA}
\affiliation{Comenius University, 842 48 Bratislava, Slovakia; Institute of Experimental Physics, 040 01 Kosice, Slovakia}
\affiliation{Joint Institute for Nuclear Research, RU-141980 Dubna, Russia}
\affiliation{Duke University, Durham, North Carolina 27708, USA}
\affiliation{Fermi National Accelerator Laboratory, Batavia, Illinois 60510, USA}
\affiliation{University of Florida, Gainesville, Florida 32611, USA}
\affiliation{Laboratori Nazionali di Frascati, Istituto Nazionale di Fisica Nucleare, I-00044 Frascati, Italy}
\affiliation{University of Geneva, CH-1211 Geneva 4, Switzerland}
\affiliation{Glasgow University, Glasgow G12 8QQ, United Kingdom}
\affiliation{Harvard University, Cambridge, Massachusetts 02138, USA}
\affiliation{Division of High Energy Physics, Department of Physics, University of Helsinki and Helsinki Institute of Physics, FIN-00014, Helsinki, Finland}
\affiliation{University of Illinois, Urbana, Illinois 61801, USA}
\affiliation{The Johns Hopkins University, Baltimore, Maryland 21218, USA}
\affiliation{Institut f\"{u}r Experimentelle Kernphysik, Karlsruhe Institute of Technology, D-76131 Karlsruhe, Germany}
\affiliation{Center for High Energy Physics: Kyungpook National University, Daegu 702-701, Korea; Seoul National University, Seoul 151-742, Korea; Sungkyunkwan University, Suwon 440-746, Korea; Korea Institute of Science and Technology Information, Daejeon 305-806, Korea; Chonnam National University, Gwangju 500-757, Korea; Chonbuk National University, Jeonju 561-756, Korea}
\affiliation{Ernest Orlando Lawrence Berkeley National Laboratory, Berkeley, California 94720, USA}
\affiliation{University of Liverpool, Liverpool L69 7ZE, United Kingdom}
\affiliation{University College London, London WC1E 6BT, United Kingdom}
\affiliation{Centro de Investigaciones Energeticas Medioambientales y Tecnologicas, E-28040 Madrid, Spain}
\affiliation{Massachusetts Institute of Technology, Cambridge, Massachusetts 02139, USA}
\affiliation{Institute of Particle Physics: McGill University, Montr\'{e}al, Qu\'{e}bec, Canada H3A~2T8; Simon Fraser University, Burnaby, British Columbia, Canada V5A~1S6; University of Toronto, Toronto, Ontario, Canada M5S~1A7; and TRIUMF, Vancouver, British Columbia, Canada V6T~2A3}
\affiliation{University of Michigan, Ann Arbor, Michigan 48109, USA}
\affiliation{Michigan State University, East Lansing, Michigan 48824, USA}
\affiliation{Institution for Theoretical and Experimental Physics, ITEP, Moscow 117259, Russia}
\affiliation{University of New Mexico, Albuquerque, New Mexico 87131, USA}
\affiliation{The Ohio State University, Columbus, Ohio 43210, USA}
\affiliation{Okayama University, Okayama 700-8530, Japan}
\affiliation{Osaka City University, Osaka 588, Japan}
\affiliation{University of Oxford, Oxford OX1 3RH, United Kingdom}
\affiliation{Istituto Nazionale di Fisica Nucleare, Sezione di Padova-Trento, $^{ff}$University of Padova, I-35131 Padova, Italy}
\affiliation{University of Pennsylvania, Philadelphia, Pennsylvania 19104, USA}
\affiliation{Istituto Nazionale di Fisica Nucleare Pisa, $^{gg}$University of Pisa, $^{hh}$University of Siena and $^{ii}$Scuola Normale Superiore, I-56127 Pisa, Italy}
\affiliation{University of Pittsburgh, Pittsburgh, Pennsylvania 15260, USA}
\affiliation{Purdue University, West Lafayette, Indiana 47907, USA}
\affiliation{University of Rochester, Rochester, New York 14627, USA}
\affiliation{The Rockefeller University, New York, New York 10065, USA}
\affiliation{Istituto Nazionale di Fisica Nucleare, Sezione di Roma 1, $^{jj}$Sapienza Universit\`{a} di Roma, I-00185 Roma, Italy}
\affiliation{Rutgers University, Piscataway, New Jersey 08855, USA}
\affiliation{Texas A\&M University, College Station, Texas 77843, USA}
\affiliation{Istituto Nazionale di Fisica Nucleare Trieste/Udine, I-34100 Trieste, $^{kk}$University of Udine, I-33100 Udine, Italy}
\affiliation{University of Tsukuba, Tsukuba, Ibaraki 305, Japan}
\affiliation{Tufts University, Medford, Massachusetts 02155, USA}
\affiliation{University of Virginia, Charlottesville, Virginia 22906, USA}
\affiliation{Waseda University, Tokyo 169, Japan}
\affiliation{Wayne State University, Detroit, Michigan 48201, USA}
\affiliation{University of Wisconsin, Madison, Wisconsin 53706, USA}
\affiliation{Yale University, New Haven, Connecticut 06520, USA}

\author{T.~Aaltonen}
\affiliation{Division of High Energy Physics, Department of Physics, University of Helsinki and Helsinki Institute of Physics, FIN-00014, Helsinki, Finland}
\author{B.~\'{A}lvarez~Gonz\'{a}lez$^z$}
\affiliation{Instituto de Fisica de Cantabria, CSIC-University of Cantabria, 39005 Santander, Spain}
\author{S.~Amerio}
\affiliation{Istituto Nazionale di Fisica Nucleare, Sezione di Padova-Trento, $^{ff}$University of Padova, I-35131 Padova, Italy}
\author{D.~Amidei}
\affiliation{University of Michigan, Ann Arbor, Michigan 48109, USA}
\author{A.~Anastassov$^x$}
\affiliation{Fermi National Accelerator Laboratory, Batavia, Illinois 60510, USA}
\author{A.~Annovi}
\affiliation{Laboratori Nazionali di Frascati, Istituto Nazionale di Fisica Nucleare, I-00044 Frascati, Italy}
\author{J.~Antos}
\affiliation{Comenius University, 842 48 Bratislava, Slovakia; Institute of Experimental Physics, 040 01 Kosice, Slovakia}
\author{G.~Apollinari}
\affiliation{Fermi National Accelerator Laboratory, Batavia, Illinois 60510, USA}
\author{J.A.~Appel}
\affiliation{Fermi National Accelerator Laboratory, Batavia, Illinois 60510, USA}
\author{T.~Arisawa}
\affiliation{Waseda University, Tokyo 169, Japan}
% ...

\collaboration{CDF Collaboration\footnote[2]{With visitors from
$^a$Istituto Nazionale di Fisica Nucleare, Sezione di Cagliari, 09042 Monserrato (Cagliari), Italy,
$^b$University of CA Irvine, Irvine, CA 92697, USA,
$^c$University of CA Santa Barbara, Santa Barbara, CA 93106, USA,
$^d$University of CA Santa Cruz, Santa Cruz, CA 95064, USA,
$^e$Institute of Physics, Academy of Sciences of the Czech Republic, Czech Republic,
$^f$CERN, CH-1211 Geneva, Switzerland,
$^g$Cornell University, Ithaca, NY 14853, USA,
$^h$University of Cyprus, Nicosia CY-1678, Cyprus,
$^i$Office of Science, U.S. Department of Energy, Washington, DC 20585, USA,
$^j$University College Dublin, Dublin 4, Ireland,
$^k$ETH, 8092 Zurich, Switzerland,
$^l$University of Fukui, Fukui City, Fukui Prefecture, Japan 910-0017,
$^m$Universidad Iberoamericana, Mexico D.F., Mexico,
$^n$University of Iowa, Iowa City, IA 52242, USA,
$^o$Kinki University, Higashi-Osaka City, Japan 577-8502,
$^p$Kansas State University, Manhattan, KS 66506, USA,
$^q$Ewha Womans University, Seoul, 120-750, Korea,
$^r$University of Manchester, Manchester M13 9PL, United Kingdom,
$^s$Queen Mary, University of London, London, E1 4NS, United Kingdom,
$^t$University of Melbourne, Victoria 3010, Australia,
$^u$Muons, Inc., Batavia, IL 60510, USA,
$^v$Nagasaki Institute of Applied Science, Nagasaki, Japan,
$^w$National Research Nuclear University, Moscow, Russia,
$^x$Northwestern University, Evanston, IL 60208, USA,
$^y$University of Notre Dame, Notre Dame, IN 46556, USA,
$^z$Universidad de Oviedo, E-33007 Oviedo, Spain,
$^{aa}$CNRS-IN2P3, Paris, F-75205 France,
$^{bb}$Texas Tech University, Lubbock, TX 79609, USA,
$^{cc}$Universidad Tecnica Federico Santa Maria, 110v Valparaiso, Chile,
$^{dd}$Yarmouk University, Irbid 211-63, Jordan.
}}
\noaffiliation
 
}

\ifthenelse{\equal{\AuthorList}{false}}{
\begin{figure}  
  \leftline{\includegraphics[scale=0.5]{cdfii_thumb_logo.eps}\hfill
    CDF/PUB/EXOTIC/CDFR/10875}
\end{figure}

\vspace*{0.5cm} 

\author{The CDF Collaboration}
\affiliation{URL http://www-cdf.fnal.gov}
}

\date{\today}

\begin{abstract}

\ifthenelse{\equal{\AuthorList}{false}}{
\vspace*{0.2cm}
}
We present a search for the standard model Higgs boson produced in association with a $Z$ boson in data collected with the  CDF~II  detector at the Tevatron, corresponding to an integrated luminosity of $9.45~\mathrm{fb^{-1}}$.
In events consistent with the decay of the Higgs boson to a bottom-quark pair and the $Z$ boson to electron or muon pairs, we set $95\%$ credibility level upper limits on the $ZH$ production cross section times the $H\rightarrow b \bar{b}$ branching ratio as a function of Higgs boson mass. 
At a Higgs boson mass of $125~\mathrm{GeV}/c^2$ we observe (expect) a limit of $7.1$ ($3.9$) times the standard model value.
\end{abstract}


\maketitle


In the standard model of particle physics (SM)~\cite{sm}, electroweak symmetry breaking~\cite{higgs} generates a fundamental scalar boson known as the Higgs boson.  Although there is strong evidence of electroweak symmetry breaking, the Higgs boson has yet to be observed.
The SM does not predict the mass of the Higgs boson, $m_H$, but the combination of precision electroweak measurements~\cite{elweak}, including recent top quark and $W$ boson mass measurements from the Tevatron~\cite{topmass,wmass}, constrains $m_H<152$~GeV/$c^2$ at the 95\% confidence level.    Direct searches at LEP2~\cite{sm-lep}, the Tevatron~\cite{tevcomb2012}, and the LHC~\cite{sm-lhc} exclude all possible masses of the SM Higgs boson at the 95\% confidence level or the 95\% credibility level (C.L.), except within the ranges 116.6 -- 119.4~GeV/$c^2$ and 122.1 -- 127~GeV/$c^2$.  A SM Higgs boson in these mass ranges would be produced in the $\sqrt{s}=1.96$ TeV $p{\bar{p}}$ collisions of the Tevatron, and have a branching fraction to $b \bar{b}$ greater than 50\%~\cite{vhtheory, stange, lhcdifferential}.  While the most sensitive searches for the SM Higgs boson at the LHC are those based on Higgs boson decays to pairs of gauge bosons, the results presented here are currently the most sensitive for a SM Higgs boson decaying to a pair of $b$ quarks.  The searches at the LHC in the four-lepton and diphoton final state offer precise measurements of the mass of the Higgs boson, while the results presented here provide information about the Higgs boson's couplings to fermions and are therefore complementary to the primary LHC search modes.  In searches for the production of a Higgs boson in association with a vector boson ({\it WH } or {\it ZH }), leptonic decays of the vector boson provide effective discrimination between the expected signal and the large, uncertain hadronic backgrounds.
Searches for $p\bar{p}\rightarrow Z(\rightarrow\ell^+\ell^-)H (\rightarrow b\bar{b})$ ($\ell = $ electron or muon~\cite{TAUS}) are among the most sensitive of the Tevatron low-mass Higgs boson searches, benefiting from low background rates and the ability to fully reconstruct both $Z$ and Higgs boson resonances.
Previous searches in this final state have been reported by the LEP2, D0, CDF, CMS, and ATLAS collaborations~\cite{sm-lep, Abazov:2010zk, PRL_41, Chatrchyan:2012ww, Aad:2012VH}. 


In this Letter, we present an updated search for \zhllbb\ events in which we expand upon the techniques of the previous CDF search and analyze data corresponding to more than twice the integrated luminosity used therein~\cite{PRL_41}.
This search introduces new multivariate $b$-jet and lepton identification techniques and updated multi-stage artificial neural network (NN) background discrimination.
This results in up to a 65\% improvement in sensitivity to a Higgs boson signal compared to the methods used in our previous search~\cite{PRL_41}.
Due to the larger data set, improved $b$-jet identification techniques that differ significantly from previously used methods, and expanded online event selection, 85\% of \zhllbb\ candidate events identified in this search were not present in the search sample used in the previous analysis~\cite{PRL_41}.

The data were collected by the upgraded CDF II detector, correspond to 9.45 fb$^{-1}$ of Tevatron \ppbar\ collisions at $\sqrt{s}$=1.96 TeV, and constitute the final CDF II data set. The CDF II detector is described in detail elsewhere~\cite{cdf}. Charged-particle trajectory (track) reconstruction and momentum determination capabilities are provided by silicon-based tracking systems surrounded by a drift chamber immersed in a $1.4$~T magnetic field~\cite{silicon, cot}.
The tracking systems are surrounded by calorimeters that provide coverage for $|\eta| < 3.6$~\cite{em, had, coordinates_zh}.  Jets are identified using a cone algorithm~\cite{Bhatti:2005ai} that combines calorimeter energy deposits to form jets with a radius of $0.4$ in $\eta$-$\phi$ space.  
External to the calorimeters, an additional system of drift chambers and scintillation counters provides muon detection for $|\eta| < 1.5$~\cite{muons}. 


CDF II records only those collision events that meet the criteria of an online event selection (trigger) system.
To maximize signal acceptance we trigger inclusively on the properties of the candidate events, using data selected by three sets of trigger algorithms~\cite{Sarah,Justin}. 
The first set consists of algorithms that require the presence of one or two electron candidates.  
The electron candidates are required to have a minimum transverse energy ($E_{T}$) of 8 to 18~GeV, depending on the specific algorithm.
The second set of trigger algorithms requires the presence of a muon candidate with a minimum transverse momentum ($p_T$) of 
18 to 22~GeV/$c$, again depending on the specific algorithm.  
Because muons deposit only a small fraction of their momentum in the calorimeter, we gain additional online efficiency by using a third set of algorithms that accept events with significant missing calorimeter transverse energy~\cite{MET_ZH}, generally above 30~GeV. 
Several of the algorithms in this set impose additional requirements on the number (typically two) and transverse energy (generally greater than 10~GeV) of jets in the event.
The combined triggers have a selection efficiency of approximately 90\% (100\%) for events within the acceptance of the CDF~II detector containing two energetic muons (electrons) and two or more jets.


Additional offline requirements are imposed on the events selected by the trigger algorithms.
Several requirements are applied to select events consistent with the decay of a $Z$ boson to either pairs of electron or pairs of muons.
Electrons and muons are selected by new NN-based algorithms optimized for efficient lepton identification~\cite{Sarah,Justin}.
The NN algorithms combine muon detector, tracking, and calorimeter information, allowing for a ~20\% increase in \zll\ acceptance compared to the selections in Ref.~\cite{PRL_41}.
We reject lepton candidates with $p_T< 10$~GeV/$c$ and require that the lepton candidate pairs have opposite electric charge when they are muons, or are electrons satisfying $|\eta| < 1.1$ for each electron~\cite{CHARGE_FAIL}. 
Events in which the reconstructed $Z$ boson has a mass of less than 76~\gevcc{} or greater than 106 \gevcc\ are rejected.
In addition to a \zll\ candidate, we require the presence of a candidate \hbb\ decay, selecting events with exactly two or three jets with $|\eta|\le$ 2.0 and an $E_T>$~25~GeV.
Jet energies include corrections for local variations in calorimeter response, the energy contribution from additional \ppbar\ interactions, and corrections specific to this analysis that assume that net missing transverse energy ($\missET$)~\cite{MET_ZH} arises predominantly from the mismeasurement of jets~\cite{Bhatti:2005ai, PRL_41}.
Events in which the combined mass of the two most energetic jets is less than 25~\gevcc\ are removed.
The resulting fractional resolution of the invariant mass of pairs of jets is estimated to be 11\%~\cite{PRL_41}.

Further event selection requires that at at least one jet in the event, referred to as a {\it b-tagged jet}, be identified as consistent with the fragmentation of a $b$ quark.  The data sample that satisfies all event selection criteria apart from the requirement of $b$-tagged jets is referred to as the \emph{PreTag} sample. 
We perform the analysis on a subset of the PreTag sample that consists of events with at least one $b$-tagged jet.
We employ a new multivariate $b$-tagging algorithm specifically designed to increase the $b$-tag efficiency and reduce the contamination of incorrectly tagged $q$ jets (\emph{q=u,s,d,g}) in CDF \hbb\ searches~\cite{Freeman:2012uf}.
For each jet containing at least one charged-particle track, the algorithm produces a scalar value in the range --1 to 1.  By comparing this value to two predetermined thresholds, the jet is classified as not tagged, \emph{loose tagged} (L), or \emph{tight tagged} (T), with all tight-tagged jets also satisfying the loose-tag definition. The thresholds defining these categories are chosen to optimize the combined expected exclusion sensitivity in simulated events.  The definition of T (L) results in a per-jet tag rate of 42\% (70\%) for jets containing the fragmentation of a $b$ quark, 9\% (27\%) for jets containing the fragmentation of a charm quark and no $b$ quark, and 0.89\% (8.9\%) for jets without the fragmentation of a $b$ or charm quark.

We form four categories of events with $b$-tagged jets.
Events with two or more jets with tight $b$ tags constitute the \emph{double-tight} (TT) category.
Events with one jet with a tight $b$ tag and one or more jets with a loose $b$ tag form the \emph{tight+loose} (TL) category.
Those with one jet with a tight $b$ tag, and no other tight or loose $b$-tagged jet make up the \emph{single tight} (Tx) category.
Events with two or more jets with loose $b$ tags comprise the \emph{double-loose} (LL) category.
If a data event satisfies more than one tag category, then the category of highest expected signal-to-background ratio is chosen, ranked  TT, TL, Tx, and LL in decreasing order.
The $b$-tagging algorithm employed in this search improves sensitivity to a $ZH$ signal by  approximately 15\% compared to the strategy used in our previous Letter~\cite{PRL_41}.

The four $b$-tag categories are subject to different systematic uncertainties, background compositions, and predicted $ZH$ content, and are therefore maintained as separate analysis channels.
We further divide events by the $Z$ boson decay (\zee\ or \zmm), and again by the number of jets in the event (two or three).
In total we form 16 exclusive channels that are simultaneously examined for $ZH$ content and jointly 
used to set upper limits on $\sigma_{ZH} \times \mathcal{B}(H\rightarrow b\bar{b})$.
In simulated signal events we find a total selection efficiency of approximately $24\%$.  


Background processes that produce two leptons and two or three jets in the final state may satisfy the above selection criteria. 
Among these, the dominant background is $Z$+jets production, nearly saturated by $Z+q\bar{q}$ before $b$-tag requirements are imposed.
After $b$ tagging, $Z+b\bar{b}$ and $Z+c\bar{c}$ are the most significant backgrounds.
$Z$+jets events are modeled using {\sc alpgen}~\cite{Mangano:2002ea} with \PYTHIA~\cite{PYTHIA} for particle showering and hadronization.
Simulated $Z$+jets samples are normalized to match experimental measurements~\cite{inclusive_kfactor_cdf} of the $Z$+jets production rate.
As reported in Refs.~\cite{CDFratio,D0ratio}, {\sc alpgen} underestimates the fraction of $Z$+heavy-flavor ($b$ and $c$) jet events in inclusive $Z$+jets production.
To compensate, we increase the normalization of $Z+b\bar{b}$ and $Z+c\bar{c}$ samples by a factor of 1.4 relative to the normalization of $Z+q\bar{q}$ samples.  

Signal, $t\bar{t}$, and diboson ($WW$, $WZ$, $ZZ$) processes are modeled with \PYTHIA.
The production rate of $ZH$ and the Higgs boson branching ratios are set to the values in Refs.~\cite{vhtheory}.
The $t\bar{t}$ simulation assumes a top-quark mass of 172.5~\gevcc{} and is normalized to a production rate of 7.04 pb~\cite{mochuwer}.
Diboson contributions are normalized to next-to-leading-order cross sections~\cite{mcfm}.
Each simulated sample includes a detailed {\sc geant}-based detector simulation~\cite{GEANT3} and uses the CTEQ5L~\cite{CTEQ5L} parton distribution functions.

We account for the contributions from QCD multijet and $W$+jets processes using a data-derived model for misidentified \zll\ candidates.
An electron and a jet have a small ($<10^{-3}$) likelihood of being misidentified as two electrons.
We model such misidentified \zee\ candidates using events containing a single electron and several jets.  
Each electron-jet pair in these events contributes to the model of misidentified \zee\, weighted by a factor reflecting the probability of the jet to be misidentified as an electron.
The determination of the weights is described in Ref.~\cite{Sarah}.
The misidentified \zmm\ contribution is modeled using like-sign muon pairs identified in the PreTag data~\cite{Justin}.


We apply several corrections that affect the normalization of simulated samples.  
We correct the instantaneous luminosity profile of the simulated samples to match that observed in data.
We correct the energy of lepton candidates to ensure agreement between the energy distributions in measured and simulated events, with corrections being approximately 1\% of the uncorrected value.
In addition, we apply corrections for differences in lepton and $b$ jet reconstruction and selection efficiencies in data and simulated samples.
To account for the selection efficiency of the CDF II trigger system, we employ multivariate trigger emulation~\cite{Justin, Sarah}.
For each of the three sets of triggers detailed above, a NN is trained on data events to describe the likelihood that the trigger system will select the event.  The training data is selected via triggers independent to the set which each seeks to describe, using the same event kinematic information as the trigger system. The output of each NN is applied to each simulated event as a normalization factor, to reflect the per-event, kinematics-dependent probability of online selection as observed in data.  
Combining all background processes, we expect a total PreTag background of $19~000 \pm 4~000$ events,
events, in good agreement with the observed total of $19~302$.
Event totals for observed data and expectations in the $b$-tagged sample are also in good agreement, with the background composition and totals listed for each $b$-tag category separately in Table~\ref{TOTALS}.


\begin{table}
\begin{tabular}{c  c c c c }
 \hline
 \hline 
 Process  &{TT} ~~& {TL} ~~& {Tx}  ~~& {LL} \\
 \hline 
$t\bar{t}$  &  55 $\pm$ 8.3 & 60 $\pm$ 8.5 & 90 $\pm$ 12 & 17 $\pm$ 2.5 \\
Diboson  &  10 $\pm$ 1.5 & 14 $\pm$ 1.9 & 40 $\pm$ 4.0 & 8.7 $\pm$ 1.0 \\
$Z+b\bar{b} $ &  59 $\pm$ 25 & 83 $\pm$ 35 & 239 $\pm$ 101 & 32 $\pm$ 14 \\
$Z+c\bar{c} $  &  3.9 $\pm$ 1.7 & 19 $\pm$ 8.4 & 109 $\pm$ 47 & 24 $\pm$ 11 \\
$Z+q\bar{q}$ &  1.0 $\pm$ 0.4 & 14 $\pm$ 3.5 & 192 $\pm$ 44 & 55 $\pm$ 14 \\
Misid. $Z$  &  2.1 $\pm$ 1.0 & 15 $\pm$ 7.6 & 31 $\pm$ 15.4 & 10 $\pm$ 5.1 \\ 
 \hline 
$ZH$ (predicted) &  1.9 $\pm$ 0.3 & 2.0 $\pm$ 0.3 & 2.8 $\pm$ 0.4 & 0.5 $\pm$ 0.1 \\
Total bkg.  &  131 $\pm$ 26 & 205 $\pm$ 38 & 701 $\pm$ 122 & 147 $\pm$ 23 \\
\hline 				
Data	&  117 &199 & 730 & 165 \\
  \hline
        \hline
         \end{tabular}
\caption{Comparison of the expected event totals for background and $ZH$ signal with the observed number of data events.  Event totals are displayed grouped by $b$-tag category (TT, TL, Tx, LL).  The $ZH$ totals assume $m_H =$~125~\gevcc.  The displayed uncertainties are systematic.  Statistical uncertainties are negligible for all model components except \mbox{misidentified $Z$}, for which they are comparable to the systematic uncertainty.  }
   \label{TOTALS}
\end{table}


To separate a possible Higgs boson signal from background, we employ a method that utilizes NN discriminants.
The multi-stage discriminant method enhances the isolation of simulated signal from background by combining a series of \emph{expert} NN's with a \emph{master} network.
The master network is constructed to isolate the $ZH$ signal from all backgrounds simultaneously, while each expert network is optimized for discrimination against a single background component.  Each NN is trained using simulated events meeting PreTag selection requirements. 
A $t\bar{t}$ expert network separates $ZH$ from $t\bar{t}$, a second $Z$+jets expert network separates signal from $Z+q\bar{q}$ and $Z+c\bar{c}$, and a third diboson expert separates $ZH$ from diboson processes.  No network specifically optimized for discriminating misidentified $Z$ events is used, because they are observed to be well separated from $ZH$ events using only the $t\bar{t}$ expert, due to their characteristically large values of $\missET$.

The final analysis is performed using the distribution of the master network scores for observed events in a binned final discriminant (BFD).
A  master network is optimized for 13 $m_H$-hypotheses ($90$ to $150~\unit{GeV}/c^2$ in $5~\unit{GeV}/c^2$ unit increments), with separate networks for two- and three-jet events.
Each master NN is constructed to return a score between 0 and 0.25 for each event, while each expert returns a value between 0 and 1, with 0 being most background-like in all cases.
The BFD has four regions (I, II, III, IV) each with a varying signal expectation and background composition. 
Events are sorted into one of the regions based on the output of the three expert networks.
If the $t\bar{t}$ expert returns a value of less than 0.5 ($t\bar{t}$-like), the event is assigned to region I.
Otherwise, if the expert for $Z+q\bar{q}$ and $Z+c\bar{c}$ returns a score of less than 0.5 ($Z+q\bar{q}$/$Z+c\bar{c}$-like), the event is assigned to region II.
Remaining events for which the diboson expert returns a value of less than 0.5 (diboson-like) are assigned to region III, with the remaining events being assigned to region IV.

\begin{figure}[t]
\begin{center}
\includegraphics[width=0.9\columnwidth]{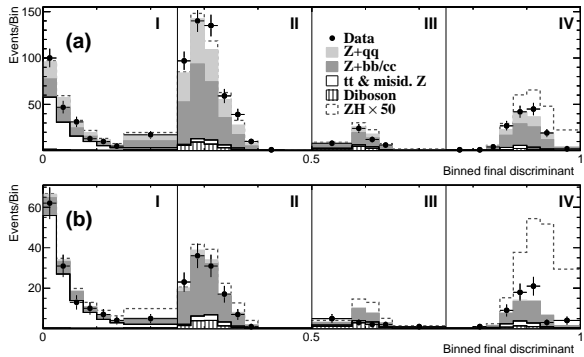}
\caption{Distribution of the BFD output for all candidates meeting Tx or LL (a) and TT or TL (b) selections, compared to the sum of the expectation from background. 
A variable bin width is used to maintain sufficient statistics in simulated samples.
The labels (I, II, III, IV) and vertical solid lines indicate the regions defined by the multi-stage discriminant method.}
\label{fig:SORT}
\end{center}
\end{figure}

The BFD is formed from the distribution of the master NN outputs plus an offset factor.
Offset factors of 0, 0.25, 0.5, and 0.75 are set for events assigned to regions I, II, III, and, IV, respectively.
The output of the BFD is shown in Fig.~\ref{fig:SORT}(a) for Tx events and for the sum of TT, TL, and LL in Fig~\ref{fig:SORT}(b) .  Histogram bins containing the highest expected ratio of signal-to-background in each region are those corresponding to higher BFD values, and the region of highest expected signal-to-background on average is region IV.
The multi-stage discriminant technique enhances sensitivity to a Higgs boson signal by approximately 10\% compared to the discriminant techniques employed in Ref.~\cite{PRL_41}.

We investigate the effect of several sources of systematic uncertainty on the search by propagating these uncertainties into the BFD distribution of the background and signal models.  The uncertainty on the measured jet energy scale (JES) is observed to significantly affect both the rate and shape of the BFD distribution.   BFD shapes generated by varying the JES by one standard deviation prior to event selection and reconstruction are used in the search for all simulated samples.  Other systematic uncertainties are found to have a negligible impact on the shape of the BFD distribution and therefore are included as uncertainties affecting process rates.
Uncertainty in the normalization of each simulated sample arises due to 
uncertainty in the integrated luminosity (6\%), 
trigger efficiency (1--5\%), 
the lepton energy scale (1.5\%), 
the amount of initial or final state radiation (1--15\%), 
$b$-tag algorithm efficiencies and $q$-jet tag probability (5--20\%), 
and the JES (5--15\%).
The JES and $b$-tag algorithm uncertainties dominate.

A 50\% uncertainty affects the normalization of the misidentified \zll\ prediction, uncorrelated between electron and muon samples.
Uncertainties of 10\%~\cite{mochuwer}, 6\%~\cite{mcfm}, 40\%, and 40\% are assumed for the normalization of top,  diboson, $Z+b\bar{b}$, and $Z+c\bar{c}$ backgrounds, respectively.
We assign a 5\% uncertainty on the normalization of $ZH$ signal samples, and account for uncertainties on the value of $\mathcal{B}(H\rightarrow b \bar{b})$~\cite{prophecy4f}.
In total, systematic uncertainties degrade sensitivity to a $ZH$ signal by approximately 13\%.

\begin{figure}[t]
\begin{center}
\includegraphics[width=0.9\columnwidth]{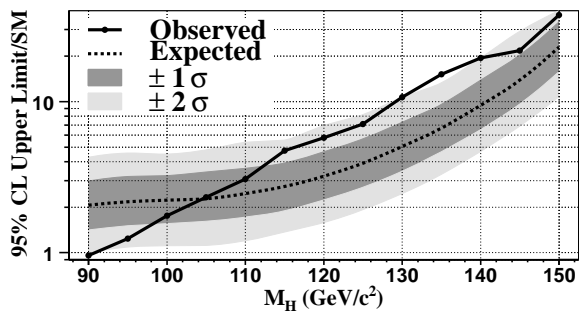}
\caption{Expected (dashed curve) and observed (solid line) $ZH$ cross section times branching fraction $95\%$ C.L. upper limits divided by the SM prediction are shown as a function of the Higgs boson mass.
The dark (light) band represents the $\pm1\sigma$ ($\pm2\sigma$) expected limit range. }

\label{fig:lim}
\end{center}
\end{figure}

We extract upper limits on the value of $\sigma_{ZH} \times \mathcal{B}(H\rightarrow b\bar{b})$ production rate using a Bayesian likelihood~\cite{PDG2012} formed as a product of likelihoods over bins of the BFD distribution for all $b$-tagged candidates.
We assume a uniform prior on the signal rate, and Gaussian priors for each systematic uncertainty, truncated so that no prediction is negative.
We set Bayesian 95\% C.L. upper limits on $\sigma_{ZH} \times \mathcal{B}(H\rightarrow b\bar{b})$ for each $m_H$ hypothesis.  Expected upper limits are derived by randomly generating a series of statistical trials, derived from the background prediction and systematic uncertainties, and computing the median of the distribution of resulting upper limits.
The upper limits on  $\sigma_{ZH} \times \mathcal{B}(H\rightarrow b\bar{b})$ are displayed in Fig.~\ref{fig:lim} and Table~\ref{table:lim2}.

We observe a broad excess for $m_H> 110~\unit{GeV}/c^2$ peaking at $135~\unit{GeV}/c^2$ with local significance of 2.4 standard deviations.  Taking the limited $m_H$ resolution of our BFD we account for a look-elsewhere effect of two, yielding a global significance of 2.1 standard deviations~\cite{Dunn1961Multiple, LEE}.

\begin{table*}[htdp]	
\begin{center}
\begin{tabular}{c c c c c c c c c c c c c c}

 \hline
 \hline
$m_H \left(\unit{GeV}/c^2\right)$ & 90 & 95 & 100 & 105 & 110 & 115 & 120 & 125 & 130 & 135 & 140 & 145 & 150 \\
\hline
Exp. & 2.1 & 2.2 & 2.2 & 2.3 & 2.5 & 2.7 & 3.2 & 3.9 & 5.0 & 6.7 & 9.4 & 13.9 & 23.0  \\
Obs. & 1.0 & 1.2 & 1.8 & 2.3 & 3.1 & 4.7 & 5.8 & 7.1 & 10.7 & 15.2 & 19.4 & 21.8 & 37.5 \\
\hline
 \hline
  \end{tabular}
 \end{center}
\caption{Expected (Exp.) and observed (Obs.) $95\%$ C.L. upper limits on the $ZH$ production cross section times the branching ratio for $H\rightarrow b \bar{b}$ normalized to the SM value for Higgs boson masses ($m_H$) between $90$ and $150~\unit{GeV}/c^2$.}
\label{table:lim2}
\end{table*}


In conclusion, we have searched for the SM Higgs boson produced in association with a $Z$ boson, followed by the decays $Z \rightarrow \ell^+ \ell^-$ and $H \rightarrow b \bar{b}$.  Finding no significant evidence for the process, we set 95\% C.L. upper limits on the $ZH$ production cross section times the $H\rightarrow b\bar{b}$ branching ratio for Higgs boson masses between $90$ and  $150\unit{GeV/}c^2$.
For a Higgs boson mass of $125\unit{GeV}/c^2$ we observe (expect) a $95\%$ C.L. upper limit of $7.1$ ($3.9$) times the standard model prediction.
Utilization of the full CDF II data set has improved sensitivity to a $ZH$ signal by 34\% compared to the previously published analysis~\cite{PRL_41}. Improved analysis methods have produced an additional approximately 30\% enhancement in sensitivity, resulting in the most sensitive search for \zhllbb\ to date.

%
We thank the Fermilab staff and the technical staffs of the
participating institutions for their vital contributions. This work
was supported by the U.S. Department of Energy and National Science
Foundation; the Italian Istituto Nazionale di Fisica Nucleare; the
Ministry of Education, Culture, Sports, Science and Technology of
Japan; the Natural Sciences and Engineering Research Council of
Canada; the National Science Council of the Republic of China; the
Swiss National Science Foundation; the A.P. Sloan Foundation; the
Bundesministerium f\"ur Bildung und Forschung, Germany; the Korean
World Class University Program, the National Research Foundation of
Korea; the Science and Technology Facilities Council and the Royal
Society, UK; the Russian Foundation for Basic Research; the Ministerio
de Ciencia e Innovaci\'{o}n, and Programa Consolider-Ingenio 2010,
Spain; the Slovak R\&D Agency; the Academy of Finland; and the
Australian Research Council (ARC).
\bibliographystyle{apsrev}

\begin{thebibliography}{42}
\expandafter\ifx\csname natexlab\endcsname\relax\def\natexlab#1{#1}\fi
\expandafter\ifx\csname bibnamefont\endcsname\relax
  \def\bibnamefont#1{#1}\fi
\expandafter\ifx\csname bibfnamefont\endcsname\relax
  \def\bibfnamefont#1{#1}\fi
\expandafter\ifx\csname citenamefont\endcsname\relax
  \def\citenamefont#1{#1}\fi
\expandafter\ifx\csname url\endcsname\relax
  \def\url#1{\texttt{#1}}\fi
\expandafter\ifx\csname urlprefix\endcsname\relax\def\urlprefix{URL }\fi
\providecommand{\bibinfo}[2]{#2}
\providecommand{\eprint}[2][]{\url{#2}}

\bibitem[{sm()}]{sm}
\bibinfo{note}{S.~Glashow, Nucl. Phys. {\bf 22}, 579 (1961); S.~Weinberg, Phys.
  Rev. Lett. {\bf 19}, 1264 (1967); A.~Salam, {\it Elementary Particle Theory},
  et. N.~Svartholm (Almquist and Wiksells, Stockholm), 367 (1968).}

\bibitem[{hig()}]{higgs}
\bibinfo{note}{F.~Englert and R.~Brout, Phys. Rev. Lett.~{\bf 13}, 321 (1964);
  P.~W.~Higgs, Phys. Rev. Lett.~{\bf 13}, 508 (1964); G.~S.~Guralnik,
  C.~R.~Hagen, and T.~W.~B.~Kibble, Phys. Rev. Lett.~{\bf 13}, 585 (1964).}

\bibitem[{elw()}]{elweak}
\bibinfo{note}{The ALEPH, CDF, D0, DELPHI, L3, OPAL, and SLD Collaboarations,
  the LEP Electroweak Working Group, the Tevatron Electroweak Working Group,
  and the SLD Electroweak and Heavy Flavour Working Groups, arXiv:1012.2367v2
  (2011).}

\bibitem[{top()}]{topmass}
\bibinfo{note}{The CDF and D0 Collaborations and the Tevatron Electroweak
  Working Group, arXiv:1207.1069 (2012).}

\bibitem[{wma()}]{wmass}
\bibinfo{note}{The CDF and D0 Collaborations and the Tevatron Electroweak
  Working Group, arXiv:1204.0042v2 (2012).}

\bibitem[{sm-({\natexlab{a}})}]{sm-lep}
\bibinfo{note}{The ALEPH, DELPHI, L3 and OPAL Collaborations, and the LEP
  Working Group for Higgs Boson Searches, Phys.\ Lett. B {\bf 565}, 61 (2003).}

\bibitem[{tev()}]{tevcomb2012}
\bibinfo{note}{The CDF and D0 Collaborations and the Tevatron New Physics and
  Higgs Working Group, arXiv:1207.0449v2 (2012).}

\bibitem[{sm-({\natexlab{b}})}]{sm-lhc}
\bibinfo{note}{S.~Chatrchyan {\it et al.} (CMS Collaboration), Phys.\ Lett.\ B
  {\bf 710}, 26 (2012); G.~Aad {\it et al.} (ATLAS Collaboration), Phys. Lett.
  B {\bf 710}, 49 (2012).}

\bibitem[{vht()}]{vhtheory}
\bibinfo{note}{J.~Baglio and A.~Djouadi, J. High Energy Phys. 10 (2010) 064;
  O.~Brein, R.~V.~Harlander, M.~Weisemann, and T.~Zirke, Eur. Phys. J. C {\bf
  72}, 1868 (2012).}

\bibitem[{sta()}]{stange}
\bibinfo{note}{A.~Stange, W.~Marciano, and S.~Willenbrock, Phys. Rev. D {\bf
  49}, 1354 (1994); A.~Stange, W.~Marciano, and S.~Willenbrock, Phys. Rev. D
  {\bf 50}, 4491 (1994).}

\bibitem[{lhc()}]{lhcdifferential}
\bibinfo{note}{S.~Dittmaier {\it et al.} (LHC Higgs Cross Section Working
  Group), arXiv:1201.3084v1 (2012).}

\bibitem[{TAU()}]{TAUS}
\bibinfo{note}{We include the small contribution from electrons and muons
  produced in tau decays ($Z \rightarrow \tau^+ \tau^- \rightarrow \ell^+
  \ell^- + X$) for both signal and background processes.}

\bibitem[{\citenamefont{Abazov et~al.}(2010)}]{Abazov:2010zk}
\bibinfo{author}{\bibfnamefont{V.~M.} \bibnamefont{Abazov}}
  \bibnamefont{et~al.} (\bibinfo{collaboration}{D0 Collaboration}),
  \bibinfo{journal}{Phys. Rev. Lett.} \textbf{\bibinfo{volume}{105}},
  \bibinfo{pages}{251801} (\bibinfo{year}{2010}).

\bibitem[{\citenamefont{Aaltonen et~al.}(2010)}]{PRL_41}
\bibinfo{author}{\bibfnamefont{T.}~\bibnamefont{Aaltonen}} \bibnamefont{et~al.}
  (\bibinfo{collaboration}{CDF Collaboration}), \bibinfo{journal}{Phys. Rev.
  Lett.} \textbf{\bibinfo{volume}{105}}, \bibinfo{pages}{251802}
  (\bibinfo{year}{2010}).

\bibitem[{\citenamefont{Chatrchyan et~al.}(2012)}]{Chatrchyan:2012ww}
\bibinfo{author}{\bibfnamefont{S.}~\bibnamefont{Chatrchyan}}
  \bibnamefont{et~al.} (\bibinfo{collaboration}{CMS Collaboration})
  (\bibinfo{year}{2012}), \eprint{CERN-PH-EP-2012-040}.

\bibitem[{\citenamefont{Aad et~al.}(2012)}]{Aad:2012VH}
\bibinfo{author}{\bibfnamefont{G.}~\bibnamefont{Aad}} \bibnamefont{et~al.}
  (\bibinfo{collaboration}{ATLAS Collaboration}) (\bibinfo{year}{2012}),
  \eprint{ATLAS-CONF-2012-015}.

\bibitem[{cdf()}]{cdf}
\bibinfo{note}{D.~Acosta, {\it et al.}, Phys. Rev. D {\bf 71}, 032001 (2005);
  D. Acosta, {\it et al.}, Phys. Rev. D {\bf 71}, 052003 (2005); A. Abulencia,
  {\it et al.}, J. Phys. G Nucl. Part. Phys. {\bf 34}, 2457 (2007).}

\bibitem[{sil()}]{silicon}
\bibinfo{note}{A.~Sill, Nucl. Instrum. Methods A {\bf 447}, 1 (2000);
  A.~Affolder {\it et al.}, Nucl. Instrum. Methods A {\bf 453}, 84 (2000);
  A.~Hill, Nucl. Instrum. Methods A {\bf 511}, 118 (2003).}

\bibitem[{cot()}]{cot}
\bibinfo{note}{A.~Affolder {\it et al.}, Nucl. Instrum. Methods A {\bf 526},
  249 (2004).}

\bibitem[{em()}]{em}
\bibinfo{note}{L.~Balka {\it et al.}, Nucl. Instrum. Methods A {\bf 267}, 272
  (1988); M.~G.~Albrow {\it et al.}, Nucl. Instrum. Methods A {\bf 480}, 524
  (2002).}

\bibitem[{had()}]{had}
\bibinfo{note}{S.~Bertolucci {\it et al.}, Nucl. Instrum. Methods A {\bf 267},
  301 (1988).}

\bibitem[{coo()}]{coordinates_zh}
\bibinfo{note}{We use a cylindrical coordinate system with $z$ along the proton
  beam direction, $r$ the perpendicular radius from the central axis of the
  detector, and $\phi$ the azimuthal angle. For $\theta$ the polar angle from
  the proton beam, we define $\eta = -\ln\tan(\theta/2)$, 
  transverse momentum $p_T = p \sin\theta$ and transverse energy $E_T = E
  \sin\theta$.}

\bibitem[{Bha()}]{Bhatti:2005ai}
\bibinfo{note}{A.~Bhatti et~al. Nucl.\ Instrum.\ Methods\ A {\bf 566}, 375
  (2006).}

\bibitem[{muo()}]{muons}
\bibinfo{note}{G.~Ascoli {\it et al.}, Nucl. Instrum. Methods A {\bf 268}, 33
  (1988).}

\bibitem[{\citenamefont{Lockwitz}(2012)}]{Sarah}
\bibinfo{author}{\bibfnamefont{S.}~\bibnamefont{Lockwitz}}
  (\bibinfo{year}{2012}), \bibinfo{note}{{Ph.D. Thesis, Yale Univ.,
  FERMILAB-THESIS-2012-02}}.

\bibitem[{\citenamefont{Pilot}(2011)}]{Justin}
\bibinfo{author}{\bibfnamefont{J.}~\bibnamefont{Pilot}} (\bibinfo{year}{2011}),
  \bibinfo{note}{{Ph.D. Thesis, The Ohio State Univ.,
  FERMILAB-THESIS-2011-42}}.

\bibitem[{MET()}]{MET_ZH}
\bibinfo{note}{The calorimter missing $E_T$ ($\missETvec(\mbox{cal})$) is
  defined by the sum over calorimeter towers, $\missETvec(\mbox{cal}) = -
  \sum_{i} E_T^i \hat{n}_i$, where $i$ is calorimeter tower number with $|\eta|
  < 3.6$, $\hat{n}_i$ is a unit vector perpendicular to the beam axis and
  pointing at the $i$th calorimeter tower. The reconstructed missing energy,
  $\missETvec$, is derived by subtracting from $\missETvec(\mbox{cal})$
  components of the event not registered by the calorimeter, such as muons and
  jet energy adjustments. $\missET(\mbox{cal})$ and $\missET$ are the scalar
  magnitudes of $\missETvec(\mbox{cal})$ and $\missETvec$, respectively.}

\bibitem[{CHA()}]{CHARGE_FAIL}
\bibinfo{note}{Electron charge determination is inaccurate for $|\eta|>$1.1 due
  to reduced tracker coverage.}

\bibitem[{\citenamefont{Freeman et~al.}(2012)\citenamefont{Freeman, Junk,
  Kirby, Oksuzian, Phillips, Snider, Trovato, Vizan, and Yao}}]{Freeman:2012uf}
\bibinfo{author}{\bibfnamefont{J.}~\bibnamefont{Freeman}},
  \bibinfo{author}{\bibfnamefont{T.}~\bibnamefont{Junk}},
  \bibinfo{author}{\bibfnamefont{M.}~\bibnamefont{Kirby}},
  \bibinfo{author}{\bibfnamefont{Y.}~\bibnamefont{Oksuzian}},
  \bibinfo{author}{\bibfnamefont{T.}~\bibnamefont{Phillips}},
  \bibinfo{author}{\bibfnamefont{F.~D.} \bibnamefont{Snider}},
  \bibinfo{author}{\bibfnamefont{M.}~\bibnamefont{Trovato}},
  \bibinfo{author}{\bibfnamefont{J.}~\bibnamefont{Vizan}}, \bibnamefont{and}
  \bibinfo{author}{\bibfnamefont{W.~M.} \bibnamefont{Yao}}
  (\bibinfo{year}{2012}), \eprint{arXiv:1205.1812}.

\bibitem[{\citenamefont{Mangano et~al.}(2003)\citenamefont{Mangano, Moretti,
  Piccinini, Pittau, and Polosa}}]{Mangano:2002ea}
\bibinfo{author}{\bibfnamefont{M.~L.} \bibnamefont{Mangano}},
  \bibinfo{author}{\bibfnamefont{M.}~\bibnamefont{Moretti}},
  \bibinfo{author}{\bibfnamefont{F.}~\bibnamefont{Piccinini}},
  \bibinfo{author}{\bibfnamefont{R.}~\bibnamefont{Pittau}}, \bibnamefont{and}
  \bibinfo{author}{\bibfnamefont{A.~D.} \bibnamefont{Polosa}},
  \bibinfo{journal}{J. High Energy. Phys.} \textbf{\bibinfo{volume}{0307}},
  \bibinfo{pages}{001} (\bibinfo{year}{2003}).

\bibitem[{PYT()}]{PYTHIA}
\bibinfo{note}{T.~Sjostrand, S.~Mrenna, and P.~Skands, J.~High Energy Phys. 05
  (2006) 026. We use \PYTHIA{} version 6.216 to generate the Higgs boson
  signals.}

\bibitem[{\citenamefont{Aaltonen et~al.}(2008)}]{inclusive_kfactor_cdf}
\bibinfo{author}{\bibfnamefont{T.}~\bibnamefont{Aaltonen}} \bibnamefont{et~al.}
  (\bibinfo{collaboration}{CDF Collaboration}), \bibinfo{journal}{Phys. Rev.
  Lett.} \textbf{\bibinfo{volume}{100}}, \bibinfo{pages}{102001}
  (\bibinfo{year}{2008}).

\bibitem[{\citenamefont{Aaltonen et~al.}(2009)}]{CDFratio}
\bibinfo{author}{\bibfnamefont{T.}~\bibnamefont{Aaltonen}} \bibnamefont{et~al.}
  (\bibinfo{collaboration}{CDF Collaboration}), \bibinfo{journal}{Phys. Rev. D}
  \textbf{\bibinfo{volume}{79}}, \bibinfo{pages}{052008}
  (\bibinfo{year}{2009}).

\bibitem[{\citenamefont{Abazov et~al.}(2011)}]{D0ratio}
\bibinfo{author}{\bibfnamefont{V.~M.} \bibnamefont{Abazov}}
  \bibnamefont{et~al.} (\bibinfo{collaboration}{D0 Collaboration}),
  \bibinfo{journal}{Phys. Rev. D} \textbf{\bibinfo{volume}{83}},
  \bibinfo{pages}{031105} (\bibinfo{year}{2011}).

\bibitem[{moc()}]{mochuwer}
\bibinfo{note}{S.~Moch and P.~Uwer, Nucl.\ Phys.\ Proc.\ Suppl.\ {\bf 183}, 75
  (2008).}

\bibitem[{mcf()}]{mcfm}
\bibinfo{note}{J.~M.~Campbell and R.~K.~Ellis, Phys.\ Rev.\ D {\bf 60}, 113006
  (1999).}

\bibitem[{GEA()}]{GEANT3}
\bibinfo{note}{GEANT, Detector description and simulation tool, CERN Program
  Library Long Writeup W5013 (1993).}

\bibitem[{CTE()}]{CTEQ5L}
\bibinfo{note}{H.~L.~Lai {\it et al.}, Eur. Phys. J. C~{\bf 12}, 375 (2000).}

\bibitem[{pro()}]{prophecy4f}
\bibinfo{note}{A.~Bredenstein, A.~Denner, S.~Dittmaier, and M.~M.~Weber, Phys.
  Rev. D {\bf 74}, 013004 (2006); A. Bredenstein, A.~Denner, S.~Dittmaier, A.
  M\H{u}ck, and M.~M.~Weber, J. High Energy Phys. 02 (2007) 080.}

\bibitem[{\citenamefont{Beringer et~al.}(2012)}]{PDG2012}
\bibinfo{author}{\bibfnamefont{J.}~\bibnamefont{Beringer}} \bibnamefont{et~al.}
  (\bibinfo{collaboration}{Particle Data Group}), \bibinfo{journal}{Phys. Rev.
  D} \textbf{\bibinfo{volume}{86}}, \bibinfo{pages}{010001}
  (\bibinfo{year}{2012}).

\bibitem[{\citenamefont{Dunn}(1961)}]{Dunn1961Multiple}
\bibinfo{author}{\bibfnamefont{O.~J.} \bibnamefont{Dunn}},
  \bibinfo{journal}{Journal of the American Statistical Association}
  \textbf{\bibinfo{volume}{56}}, \bibinfo{pages}{52} (\bibinfo{year}{1961}).

\bibitem[{LEE()}]{LEE}
\bibinfo{note}{L.~Lyons, The Annals of Applied Statistics, Vol. 2, No. 3, 887
  (2008).}

\end{thebibliography}


\end{document}